\documentclass[twocolumn]{jpsj3}

\usepackage{color}
\usepackage{graphicx}
\usepackage{dcolumn}
\usepackage{bm}

\title{Ferromagnetic Quantum Critical Point Induced by Tuning the Magnetic Dimensionality of the Heavy-Fermion Iron Oxypnictide Ce(Ru$_{1-x}$Fe$_x$)PO}

\author{Shunsaku~Kitagawa$^1$\thanks{E-mail address: shunsaku@scphys.kyoto-u.ac.jp}, Kenji~Ishida$^1$, Tetsuro~Nakamura$^2$, Masanori~Matoba$^2$, and Yoichi~Kamihara$^2$}
\inst{$^1$Department of Physics, Graduate School of Science, Kyoto University, Kyoto 606-8502, Japan \\
$^2$Departments of Applied Physics and Physico-Informatics, Keio University, Kanagawa 223-8522, Japan}

\abst{We have performed $^{31}$P-NMR measurements of the $c$-axis-aligned poly-crystal Ce(Ru$_{1-x}$Fe$_{x}$)PO with a two-dimensional layered structure in order to understand the origin of $T_{\rm Curie}$ suppression by Fe substitution. 
The variation in the magnetic-fluctuation character with respect to $x$ is investigated from the in-plane and out-of plane fluctuations and the relationship between the static spin susceptibility and the in-plane fluctuation.
It was found that three-dimensional ferromagnetic (FM) correlations are dominant and give rise to the FM ordering in CeRuPO. The out-of plane fluctuations are significantly suppressed with increasing $x$, and it was revealed that the FM fluctuations become two dimensional near a FM quantum critical point (QCP).
Our NMR results strongly suggest that a unique FM QCP in Ce(Ru$_{1-x}$Fe$_{x}$)PO is induced by the suppression of the magnetic correlations along the $c$-axis, which is a different mechanism from that presumed in heavy-fermion compounds.}

\kword{NMR, heavy fermion, quantum criticality, ferromagnetism, dimensional tuning}

\begin{document}
\maketitle

Heavy-fermion (HF) systems, in which high-temperature localized $f$ electrons behave as itinerant electrons at low temperatures, are one of the strongest electron correlation systems since they have extremely large effective electron masses\cite{T.Fujita_JMMM_1985,G.R.Stewart_RMP_1984}. 
This implies that the energy scale of the systems becomes extremely small, and thus their ground state can be tuned from the magnetically ordered state to the disordered state, and quantum criticality is induced relatively easily by controlling nonthermal external parameters\cite{A.Schroder_Nature_2000,P.Gegenwart_PRL_2002}.  
Quantum critical behavior in HF compounds has been considered to be induced by the competition of two effects: 
the Ruderman-Kittel-Kasuya-Yosida (RKKY) interaction, which is the intersite exchange interaction between localized $f$-electron moments and induces magnetic ordering, and the Kondo interaction, where localized $f$-electron moments are screened by conduction electrons. 
It forms a nonmagnetic HF state. 
Although the two interactions lead to opposite ground states, crucial parameters for the two interactions are the coupling between conduction electrons and $f$-electrons, $J_{cf}$, and the density of states around the Fermi energy $D(E_{\rm F})$; thus, the quantum critical point (QCP) in HF compounds has been presumed to be induced by tuning the strength of $J_{cf}D(E_{\rm F})$.

Up to now, there have been considerable efforts for understanding the nature of a QCP in various antiferromagnetic HF compounds, but very few in ferromagnetic (FM) HF compounds, particularly Ce-based FM HF compounds.
In our previous paper, we reported that the continuous suppression of a FM transition in Ce(Ru$_{1-x}$Fe$_{x}$)PO with an isovalent Fe substitution for Ru, and that the novel FM QCP in Ce(Ru$_{1-x}$Fe$_{x}$)PO is present at $x \sim 0.86$\cite{S.Kitagawa_PRL_2012}.
The observed criticality is in sharp contrast to the FM criticality observed in other FM compounds\cite{H.Kotegawa_JPSJ_2011,D.Aoki_JPSJ_2011}. 
For example, the FM transition in UGe$_2$ is gradually suppressed by applying pressure, but the transition changes from the second order to the first order at a tricritical point and the first-order metamagnetic transition emerges in a small external field\cite{H.Kotegawa_JPSJ_2011}. 
Similar phenomena have been observed in ZrZn$_2$\cite{M.Uhlarz_PRL_2004} and MnSi\cite{C.Pfleiderer_PRB_1997}, which are supported theoretically\cite{D.Belitz_PRL_2005,H.Yamada_PRB_1993}. 
The nature of the quantum phase transition from the FM state to the paramagnetic (PM) state has been revisited and has attracted much interest. 
Here, we show from the comprehensive $^{31}$P-NMR measurements that the FM QCP in Ce(Ru$_{1-x}$Fe$_{x}$)PO are induced by the suppression of the out-of-plane ($c$-axis) magnetic correlations, indicative of the tuning of the dimensionality of the magnetic correlations from three-dimensionality to two-dimensionality. 
This is a new route for inducing quantum criticality in not only HF compounds but also itinerant FM compounds, similarly to the case realized in superlattice HF systems\cite{H.Shishido_Science_2010}. 

\begin{figure*}[!tb]
\vspace*{0pt}
\begin{center}
\includegraphics[width=15.5cm,clip]{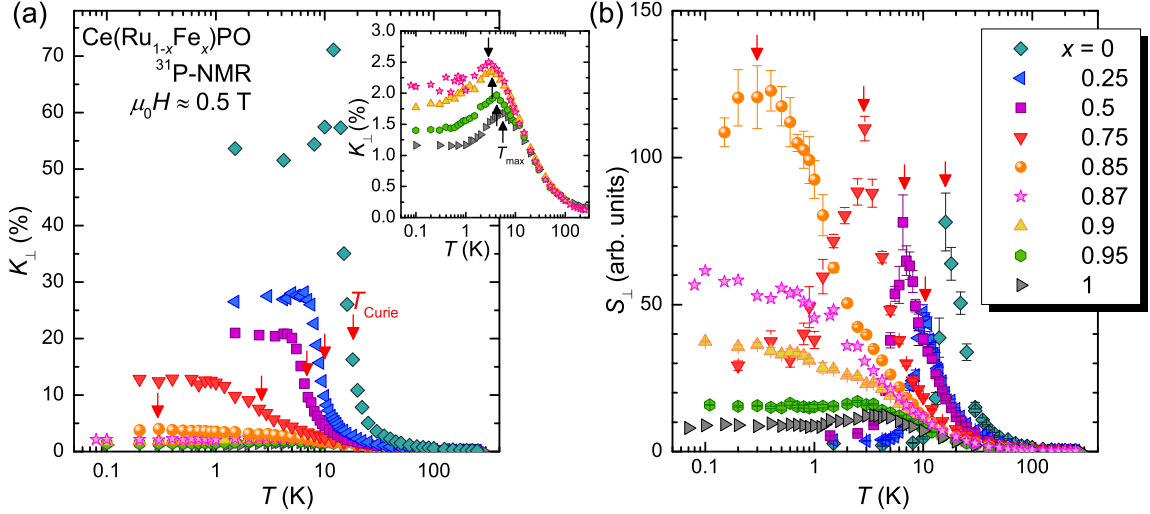}
\end{center}
\caption{(Color online) $T$ dependencess of (a) in-plane Knight shift $K_{\perp}$ and (b) low-energy in-plane spin fluctuations $S_{\perp}$ at $\mu_0H \simeq 0.5$~T for Ce(Ru$_{1-x}$Fe$_{x}$)PO. 
The date above $x$ = 0.25 has already been shown in a previous paper\cite{S.Kitagawa_PRL_2012}.
$K_{\perp}$ jumps up and $S_{\perp}$ has a peak at $T_{\rm Curie}$. 
$T_{\rm Curie}$ is determined from the peak of $S_{\perp}$. 
$T_{\rm Curie}$ and the internal field at the P site, proportional to ordered magnetic moment $\langle \mu_{\rm ord} \rangle$, continuously approach zero toward $x \sim 0.86$ with the substitution of Fe for Ru.
These results suggest the presence of FM QCP at $x \sim 0.86$. 
Above $x = 0.9$, the spin fluctuation $S_{\perp}$ as well as $K_{\perp}$ becomes almost constant at low temperatures, indicative of the formation of the PM HF state.
(Inset) The $T$ dependence of $K_{\perp}$ above $x = 0.87$.
$K_{\perp}$ shows a peak at approximately $T_{\rm max} \sim$ 5 K (indicated by the arrows) instead of the FM order, suggestive of the PM HF ground state at low fields.}
\label{Fig.1}
\vspace*{-20pt}
\end{figure*}

The iron oxypnictide CeFe(Ru)PO is a material related to the iron-based superconductor LaFePO. 
Both materials possess the same two-dimensional (2D) layered structure, stacking the Ce(La)O and Fe(Ru)P layers alternatively. 
The CeO layer contributes to the large magnetic response, and the Fe(Ru)P layer is conductive in CeFe(Ru)PO. 
CeRuPO is a FM HF system with a Curie temperature $T_{\rm Curie}$ = 15~K and a coherent temperature $T_{\rm K} \simeq 10$~K\cite{C.Krellner_PRB_2007}. 
On the other hand, the counterpart CeFePO is a HF compound with $T_{\rm K} \simeq 10$~K; its PM ground state is PM down to 80~mK\cite{E.Buning_PRL_2008,S.Kitagawa_PRL_2011}. 
Therefore, we expect that this ground state can be tuned continuously from the FM state to the PM state by substituting isovalent Fe for Ru.


Polycrystalline Ce(Ru$_{1-x}$Fe$_{x}$)PO compounds ($x = 0$, 0.25, 0.5, 0.75, 0.85, 0.87, 0.9, 0.95, and 1) was synthesized by a solid-state reaction\cite{Y.Kamihara_JPCS_2008}. 
As reported in previous papers\cite{S.Kitagawa_PRL_2011,S.Kitagawa_PRL_2012}, these compounds possess a 2D anisotropy in magnetic susceptibility, so that the polycrystalline samples were uniaxially aligned by taking advantage of such anisotropy. 
$^{31}$P-NMR measurement was performed on the $c$-axis aligned samples.

NMR measurement can probe static and dynamic magnetic properties. 
The Knight shift $K_i$($T$, $H$) ($i = \perp$ and $c$) is defined as  
\begin{equation}
K_i(T, H) = \left(\frac{H_0 - H_{\rm res}}{H_{\rm res}}\right)_{\omega = \omega_0} \propto \frac{M_i(T,H_{\rm res})}{H_{\rm res}},
\end{equation} 
where $H_{\rm res}$ is the magnetic field at the resonance peak, and  $H_0$ and $\omega_0$ are the resonance field and frequency of a bare $^{31}$P nucleus, respectively, and have the relation $\omega_0 = \gamma_n H_0$ with the $^{31}$P-nuclear gyromagnetic ratio $\gamma_n$. 
On the other hand, $1/T_1$ can be used to probe spin fluctuations perpendicular to the applied magnetic field, as discussed in a previous paper\cite{S.Kitagawa_PRL_2011}. 
Therefore, $1/T_1$ in $H \parallel c$ and $H \perp c$ are described as
\begin{align}
\left(\frac{1}{T_1}\right)_{H\parallel c} \equiv 2S_{\perp}, 
\left(\frac{1}{T_1}\right)_{H\perp c} \equiv S_c + S_{\perp}.
\end{align}
From the two equations, we can determine the low-energy $q$-summed spin fluctuation of the in-plane ($S_{\perp}$) and out-of-plane ($c$-axis) ($S_c$) components, separately.

Figure~\ref{Fig.1} shows temperature dependences of (a) $K_{\perp}$ and (b) $S_{\perp}$ for Ce(Ru$_{1-x}$Fe$_{x}$)PO ($x = 0$, 0.25, 0.5, 0.75, 0.85, 0.87, 0.9, 0.95, and 1). 
As reported in a previous paper\cite{S.Kitagawa_PRL_2012}, both $K_{\perp}$ and $S_{\perp}$ suggest the existence of the FM QCP at $x \sim 0.86$: $T_{\rm Curie}$ and the internal field at the P site, proportional to ordered magnetic moment $\langle \mu_{\rm ord} \rangle$, continuously approach zero toward $x \sim 0.86$ with the Fe substitution for Ru. 
At $x = 0.87$, spin fluctuations are continuously enhanced on cooling down to 70~mK, suggesting that the system is very close to the QCP. 
Above $x = 0.9$, $S_{\perp}$ and $K_{\perp}$ become almost constant at low temperatures, indicative of the formation of the PM HF state. 
These results indicate that the ground state continuously changes from the FM state to the PM state with the Fe substitution.


Here, we consider the effect of the Fe substitution. 
In HF compounds, a ground state changes from the magnetically ordered state to the nonmagnetic HF state with respect to the strength of $J_{cf}D(E_{\rm F})$, the so-called Doniach scenario\cite{S.Doniach_PhysicaBC_1977}. 
In the case of Ce(Ru$_{1-x}$Fe$_{x}$)PO, it is considered that the Fe substitution corresponds to the application of chemical pressure since the ion radius of Fe is smaller than that of Ru. 
This is actually observed in the $x$ dependence of the unit cell volume in Ce(Ru$_{1-x}$Fe$_{x}$)PO\cite{S.Kitagawa_PRL_2012}. 
In this case, it is expected that $J_{cf}D(E_{\rm F})$ becomes stronger with increasing $x$, resulting in a higher coherent temperature\cite{S.Kawasaki_PRB_2008_2} and the suppression of FM ordering. 
However, the coherent temperature is reported to be unchanged in CeRuPO and CeFePO\cite{C.Krellner_PRB_2007,E.Buning_PRL_2008}, indicating that the above scenario is not valid in Ce(Ru$_{1-x}$Fe$_{x}$)PO. 
To understand the origin of $T_{\rm Curie}$ suppression by the Fe substitution, we investigate the variation in magnetic-fluctuation character in Ce(Ru$_{1-x}$Fe$_{x}$)PO from the temperature dependences of $S_{\perp}$ and $S_c$ and  the relationship between $K_{\perp}$ and $(1/T_1T)_{H\parallel c}$.

\begin{figure}[tb]
\vspace*{-0pt}
\begin{center}
\includegraphics[width=8cm,clip]{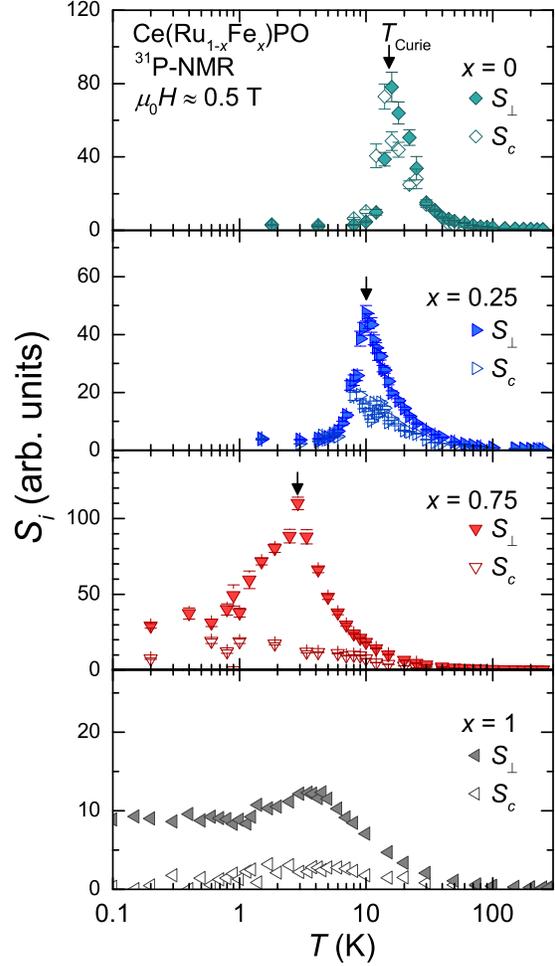}
\end{center}
\caption{(Color online)  $T$ dependences of $S_{i}$ ($i = \perp, c$) at $x = 0$, 0.25, 0.75, and 1. Although $S_{\perp}$ gradually changes with respect to $x$, $S_c$ is rapidly suppressed, suggestive of the strong suppression of the out-of-plane ($c$-axis) magnetic correlations against the Fe substitution.}
\label{Fig.2}
\end{figure}

Figure \ref{Fig.2} shows the temperature dependences of $S_{\perp}$ and $S_c$ in Ce(Ru$_{1-x}$Fe$_x$)PO with $x = 0$, 0.25, 0.75, and 1.
$S_c$ is comparable to $S_{\perp}$ and shows a clear peak at the $T_{\rm Curie}$ in CeRuPO ($x = 0$).
This indicates that the magnetic fluctuations are isotropic although the static spin susceptibility possesses the $XY$ spin anisotropy above $T_{\rm Curie}$.  
In contrast, $S_{c}$ is rapidly suppressed with increasing $x$ and no anomaly in $S_{c}$ is observed at $T_{\rm Curie}$ of the $x = 0.75$ sample, showing that the magnetic fluctuations also possess the $XY$ spin anisotropy near the FM QCP.   

Next, we discuss the magnetic-correlation character in $S_{\perp}$, predominant near $T_{\rm Curie}$.
Magnetic-correlation characters near the magnetic instability have been interpreted using the self-consistently renormalization (SCR) theory in HF compounds as well as in itinerant magnets\cite{Moriya_SCR,T.Moriya_AP_2000}. 
In this theory, $1/T_1T$ is determined by the predominant low-energy spin fluctuation in $q$-space and the dimensionality of magnetic correlations. 
Thus, when FM correlations are dominant, $1/T_1T$ is proportional to the power law of static spin susceptibility $\chi(\bm{q} = 0)$, which is equal to the Knight shift $K$ with respect to the dimensionality of the magnetic correlations: 
\begin{align*}
1/T_1T &\propto 
\begin{cases}
\chi(\bm{q} = 0) \propto K \text{~(3D FM correlations),}\\
\chi^{1.5}(\bm{q} = 0) \propto K^{1.5} \text{~(2D FM correlations).}
\end{cases}
\end{align*}
Thus, the relationship between $1/T_1T$ and $K$ gives information on the $q$-space and the dimensionality of magnetic correlations. 

\begin{figure}[!tb]
\vspace*{-0pt}
\begin{center}
\includegraphics[width=8.5cm,clip]{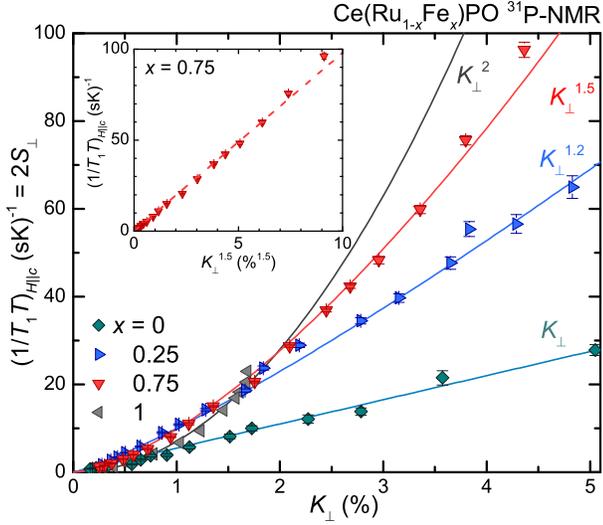}
\end{center}
\caption{(Color online) $(1/T_1T)_{H\parallel c}$ vs $K_{\perp}$ for $x = 0$, 0.25, 0.75, 1.
$(1/T_1T)_{H\parallel c} \propto S_{\perp}$ is proportional to the power law of $K_{\perp}$, which is almost equal to the static spin susceptibility $\chi(\bm{q} = 0)$ [$(1/T_1T)_{H\parallel c}$ can be described as $(1/T_1T)_{H\parallel c}$ = $AK^n$ ].
The dimensionality of the magnetic correlations can be determined from $n$ on the basis of the SCR theory. 
At $x = 0$, $n \simeq 1$ indicates that the 3D FM correlations are dominant. 
The power law of $K_{\perp}$ becomes larger with increasing $x$ and $n \simeq 1.5$ at $x = 0.75$, indicating that the FM correlations become 2D. 
With further substitution of Fe, the system becomes far from the FM QCP and $n \simeq 2$ at $x = 1$, indicative of the lack of critical fluctuations. 
These results suggest that the dimensionality of the magnetic correlations changes from 3D to 2D with the Fe substitution for Ru. 
Solid curves are visual guides. 
(Inset) $(1/T_1T)_{H\parallel c}$ versus $K^{1.5}$ at $x = 0.75$. 
A linear relation is clearly observed. 
The dashed line is a visual guide.}
\label{Fig.3}
\end{figure}

Figure~\ref{Fig.3} shows $K_{\perp}$ vs $(1/T_1T)_{H \parallel c} \propto S_{\perp}$ for $x = 0$, 0.25, 0.75, and 1. 
For all samples, $(1/T_1T)_{H \parallel c}$ is proportional to the power law of $K$ [$(1/T_1T)_{H \parallel c}$ can be described as $(1/T_1T)_{H \parallel c} = AK^n$, where $A$ is a coefficient]. 
At CeRuPO, $n \simeq 1$ indicates that three-dimensional (3D) FM correlations are dominant and give rise to FM ordering. 
The power law of $K_{\perp}$ becomes larger with increasing $x$, and $n$ is nearly 1.5 at $x = 0.75$ close to the FM QCP, indicating that the FM correlations become 2D. 
With further substitution of Fe, the system goes away from the FM QCP and $n \simeq 2$ at CeFePO ($x = 1$), indicative of the lack of critical FM fluctuations and formation of the 2D HF state.
Figures~\ref{Fig.2} and \ref{Fig.3} clearly show that FM fluctuations in the samples near the critical concentration possess 2D anisotropy in spin and $k$ spaces, respectively.
In addition, the close relationship between the $c$-axis magnetic correlations and the $c$-axis spin component is suggested by the experimental fact that a spontaneous ordered moment points to the $c$-axis although the spin-space anisotropy above $T_{\rm Curie}$ has the $XY$-type character in CeRuPO\cite{C.Krellner_JCG_2008}.  

\begin{figure}[tb]
\vspace*{-0pt}
\begin{center}
\includegraphics[width=8cm,clip]{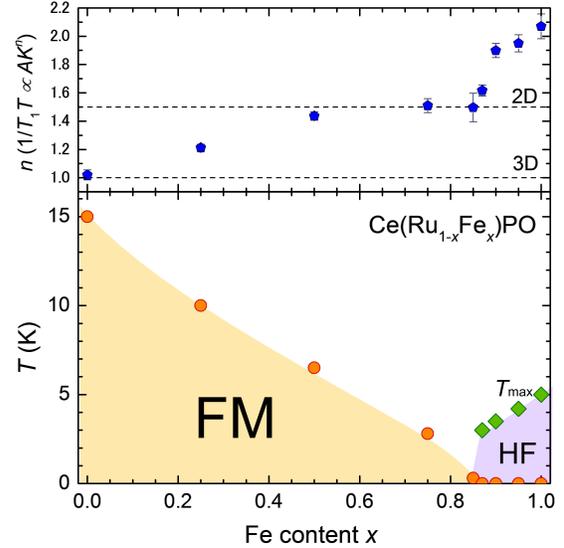}
\end{center}
\caption{(Color online) $T-x$ phase diagram for Ce(Ru$_{1-x}$Fe$_{x}$)PO. 
Filled circles represent $T_{\rm Curie}$, and filled diamonds correspond to the temperature where the Knight shift shows a peak $T_{\rm max}$. 
The power law $n$ against $x$ is estimated from the fitting between $\sim T_{\rm Curie}$ and 100~K and plotted in the above panel. 
When the dimensionality of magnetic correlations becomes 2D ($n$ becomes 1.5), $T_{\rm Curie}$ is suppressed and the FM QCP appears at $x \sim 0.86$, suggesting that the FM QCP is induced by the tuning of the dimensionality of the magnetic correlations.}
\label{Fig.4}
\end{figure}

\begin{figure}[tb]
\vspace*{-0pt}
\begin{center}
\includegraphics[width=8.8cm,clip]{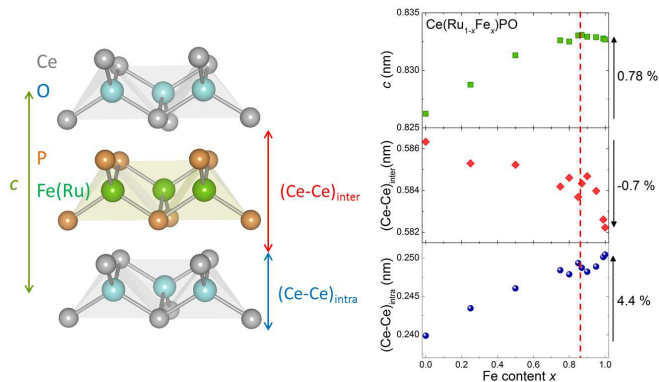}
\end{center}
\caption{(Color online) Crystal structure of Ce(Ru$_{1-x}$Fe$_{x}$)PO and Fe content $x$ dependence of the $c$-axis length, the distance between the CeO block layer along the $c$ axis (Ce-Ce)$_{\rm inter}$, and the distance between the Ce atoms in the CeO block layer (Ce-Ce)$_{\rm intra}$. 
The arrows in the left figure indicate the position of each length. 
The $c$-axis and (Ce-Ce)$_{\rm intra}$ increase with increasing $x$, although (Ce-Ce)$_{\rm inter}$ decreases. (Ce-Ce)$_{\rm intra}$ changes markedly with the Fe substitution, indicating that (Ce-Ce)$_{\rm intra}$ is the most important parameter to determine the dimensionality of the  magnetic correlations in Ce(Ru$_{1-x}$Fe$_{x}$)PO. 
The dashed line of the right figure corresponds the concentration of the FM QCP. }
\label{Fig.5}
\end{figure}
Figure~\ref{Fig.4} shows the $T-x$ phase diagram obtained from our NMR measurements. 
The $x$ dependence of the power law $n$ estimated from fitting between $\sim T_{\rm Curie}$ and 100~K indicates that the magnetic correlation along the $c$-axis becomes weaker with increasing $x$, and that FM ordering disappears with a decrease in the dimensionality of the $c$-axis correlations. 

This scenario is consistent with the Rietveld analyses of X-ray diffraction, as shown in Fig.~\ref{Fig.5}. 
The Fe substitution increases the $c$-axis lattice parameter, particularly the distance between Ce atoms along the $c$-axis in the CeO block layer, (Ce-Ce)$_{\rm intra}$. Therefore, (Ce-Ce)$_{\rm intra}$ is considered to be a crucial parameter for FM ordering. 
Taking into account all results, it is concluded that the suppression of $T_{\rm Curie}$ by Fe substitution is triggered by the suppression of the $c$-axis magnetic correlations.
According to the Mermin-Wagner theorem\cite{N.D.Mermin_PRL_1966}, electronic spins in ideal 2D systems cannot order when they hold a continuous rotational symmetry. 
Although Ce(Ru$_{1-x}$Fe$_{x}$)PO cannot be regarded as the ideal 2D system, Ce(Ru$_{1-x}$Fe$_{x}$)PO is a rare example where $T_{\rm Curie}$ is suppressed by the tuning of the dimensionality of the magnetic correlations. 

In summary, we have performed $^{31}$P-NMR measurements on Ce(Ru$_{1-x}$Fe$_{x}$)PO to investigate the $x$ dependence of magnetic correlations. 
We found that the FM QCP in Ce(Ru$_{1-x}$Fe$_{x}$)PO is induced by the suppression of the $c$-axis correlations, i.e., the tuning of the dimensionality of the magnetic correlations. 
This mechanism is quite different from that presumed in HF compounds and itinerant FM compounds, but similar to the dimensional tuning realized in superlattice CeIn$_3$/LaIn$_3$\cite{H.Shishido_Science_2010}. 
Thus, CeFePO is regarded as an ideal 2D HF compound.

\section*{Acknowledgments}
The authors thank S. Yonezawa and Y. Maeno for experimental support and valuable discussions. 
The authors are also grateful to H. Ikeda, Y. Takahashi, K. Deguchi, N. K. Sato, C. Geibel, and C. Krellner for fruitful discussions. 
This work was partially supported by Kyoto University LTM Center, a ``Heavy Electrons'' Grant-in-Aid for Scientific Research on Innovative Areas (No. 20102006) from The Ministry of Education, Culture, Sports, Science, and Technology (MEXT) of Japan, a Grant-in-Aid for the Global COE Program ``The Next Generation of Physics, Spun from Universality and Emergence'' from MEXT of Japan, a Grant-in-Aid for Scientific Research from Japan Society for Promotion of Science (JSPS), KAKENHI (S \& A) (Nos. 20224008 and 23244075), and the Funding Program for World-Leading Innovative R\&D on Science and Technology (FIRST) from JSPS. One of the authors (SK) was financially supported by a JSPS Research Fellowship.


\end{document}